\documentstyle{article}

\newcommand {\tsub}[1]{_{\mbox{\protect\scriptsize #1}}}
\newcommand {\tsup}[1]{^{\mbox{\protect\scriptsize #1}}}
\newcommand {\Ref}[1]{(\ref{#1})}
\newcommand 
{\Stretch}[1]{\renewcommand{\baselinestretch}{#1}\large\normalsize}

\newcommand {\ket}[1]{\left|#1\right\rangle}
\newcommand {\bra}[1]{\left\langle#1\right|}
\newcommand {\ketd}[1]{\left.\left|#1\right\rangle\right\rangle}
\newcommand {\brad}[1]{\left\langle\left\langle#1\right|\right.}
\newcommand {\kett}[1]{\left.\left.\left|#1\right\rangle
                                \right\rangle\right\rangle}

\begin{document}

\pagestyle{empty}

\Stretch{3}
\begin{center}

{\large 27/8/96}
\vspace*{2em}

{\huge \bf Applications of the density matrix renormalisation group to 
problems in magnetism}
\vspace*{5em}

\Stretch{1.5}
{\large
G.\ A.\ Gehring$^{1}$, R.\ J.\ Bursill$^{1\dagger}$ and T. Xiang$^{2}$
}
\vspace*{2em}

$^{1}$Department of Physics,
The University of Sheffield,
S3 7RH, Sheffield, United Kingdom.\\

$^{2}$Interdisciplinary Research Centre in Superconductivity,
The University of Cambridge, Cambridge, CB3 0HE, UK.\\

\medskip

$\dagger$ Present address: School of Physics, UNSW, Sydney, 2052,
Australia.

\end{center}

\vfill
\eject

\Stretch{1.3}

\begin{abstract}

The application of real space renormalisation group methods to quantum 
lattice models has become a topic of great interest following the 
development of the Density Matrix Renormalisation Group (DMRG) by White. 
This method has been used to find the ground and low-lying excited state 
energies and wave functions of quantum spin models in which the form of 
the ground state is not clear, for instance because the interactions are 
frustrated. It has also been applied to fermion problems where the 
tendency for localisation due to the strong Coulomb repulsion is opposed 
by the lowering of the kinetic energy which occurs as a result of 
electron transfer. The approach is particularly suitable for one, or 
quasi-one dimensional problems.

The method involves truncating the Hilbert space in a systematic and 
optimised manner. Results for the ground state energy are thus 
variational bounds. The results for low-lying energies and correlation 
functions for one dimensional systems have unprecedented accuracy and 
the method has become the method of choice for solving one dimensional 
quantum spin problems.

We review the method and results obtained for the spin-1 chain 
with biquadratic exchange as well as the spin-1/2 model with competing 
nearest and next nearest neighbour exchange will be described. More 
recently, the DMRG has been applied to reformulate the coupling constant 
renormalisation group approach which is appropriate for the study of 
critical properties. This approach has been applied to the anisotropic 
spin-1/2 Heisenberg chain.

Finally, we discuss recent work which has borne promising applications in 
two dimensions---the Ising model and the two dimensional Hubbard model.

\end{abstract}

\setcounter{page}{0}
\vfill
\eject

\pagestyle{plain}

\section{Introduction}
\label{Introduction}
\setcounter{equation}{0}

Recent times have seen the fabrication of an increasingly rich variety of 
materials whose low energy magnetic properties require intrinsically 
quantum mechanical as well as {\em low dimensional} lattice models for a 
realistic description. On the other hand, the 
development of sophisticated field theories and the surprising conjecture 
of Haldane \cite{haldane} has dramatically rekindled theoretical and 
mathematical interest in one dimensional quantum spin models. Also, the 
advent of high temperature superconductors has 
generated intensive interest in the magnetic properties of two dimensional 
quantum lattice models such as the Hubbard, $t$-$J$ and Heisenberg models.

It is well known that analytical solutions of these models are only 
available in rare and special cases and that mean field theories perform 
unreliably when applied to low-dimensional systems. Also, the presence, for 
example, of frustration in these systems renders the construction of 
suitable variational wave functions very difficult as the structure of 
the ground state is not at all clear ahead of time. As a result, 
perturbative attempts at systematically improving upon variational model 
states are often prone to slow or even spurious convergence. This problem 
of frustration has also for a long time plagued attempts at applying Monte 
Carlo methods to many of the abovementioned systems at low temperatures. 
Exact diagonalisation approaches may fail to reach 
suitably large lattices for two dimensional or coupled chain problems due 
to the exponential growth in the size of the Hilbert space. This is 
especially problematic for frustrated systems that possess incommensurate 
order in the ground state.

There has therefore been a real window of opportunity open for the 
development of a portable, robust and systematically improvable 
numerical scheme for solving low dimensional quantum lattice models of 
magnetic systems. The advent of the renormalisation group (RG) 
\cite{wilson} and Wilson's highly successful numerical solution of the 
Kondo problem \cite{wilson} in addition to promising applications to 
classical systems in two dimensions \cite{classical} suggested that the 
Real Space Renormalisation Group (RSRG) might well fill this r\^{o}le.

Unfortunately, attempts at carrying out this programme produced results 
with disappointingly slow convergence \cite{bray_chui}. The underlying 
causes 
for the poor performance of the RSRG were addressed in a seminal series 
of papers by White \cite{development} where a powerful 
improvement on the RSRG, the Density Matrix Renormalisation Group 
(DMRG), was formulated. These papers were followed with a phenomenally 
successful study of the spin-1 chain where the Haldane conjecture, that 
spin chains of integral spin are gapped, was verified with unprecedented 
accuracy \cite{s=1_chain}. A flood of applications to problems in low 
dimensional quantum magnetism has since followed.

In this article we review the DMRG and some its applications to problems 
in magnetism. In section~\ref{method} we briefly outline the method and 
the quantities of interest which can be calculated. In 
section~\ref{frustration} we discuss applications of the DMRG to spin 
chains with frustration. In section~\ref{critical} we review attempts at 
using the DMRG to study phase transitions in these systems, either by 
direct calculation of order parameters or by deriving RG transformations in 
the space of coupling constants. In section~\ref{current} we discuss 
recent extensions of the DMRG to two dimensional 
problems and one dimensional models at non-zero temperature.

\section{The DMRG method}
\label{method}

The DMRG is an iterative, truncated basis procedure whereby a large 
system (or superblock) is built up from a small number of lattice sites 
by adding a few sites at a time. As mentioned, the DRMG was introduced 
and formally developed in a series of papers \cite{development} by White 
and co-workers at Irving. We refer the reader to these excellent papers 
for a detailed description of the method. Here we will attempt to 
present a simple exposition which captures the essential 
algorithmic steps involved. It will be useful to work with a concrete 
model; the one-dimensional, spin-1/2 $XY$ model
\begin{equation}
{\cal H}
=
\sum_{i}
\left( S_{i}^{+}S_{i+1}^{-} + \mbox{ h.c.} \right)
\label{HXY}
\end{equation}
where $S_{i}^{\pm}$ is the spin-1/2 raising (lowering) operator for 
lattice site $i$.

Let us initially define a {\em system} block $A$ which consists of only 
a single site and hence has $M=2$ states $\ket{1}=\ket{\uparrow}$ and 
$\ket{2}=\ket{\downarrow}$. We record the matrix elements of the spin 
operator $S^{+}_{A}$ for this site, as well as the block Hamiltonian
${\cal H}_{A}$, with respect to the system block basis 
$\{\ket{n}:n=1,\ldots,M\}$ 
i.e.
$S(n',n)\equiv\bra{n'}S^{+}_{A}\ket{n}$ and
$H(n',n)\equiv\bra{n'}{\cal H}_{A}\ket{n}\equiv 0$.

In the first iteration we form an {\em environment} block $B$ which is 
precisely the same as the system block (a single site). Combining the 
system block, the environment block and two extra sites, we form a {\em 
superblock} $A\bullet\bullet B$. A basis for the superblock is given by
\begin{equation}
\left\{
\ket{n\tsub{S}}
\equiv
\ket{n_{A}}\otimes\ket{s_{1}}\otimes\ket{s_{2}}\otimes\ket{n_{B}}
:n_{A},n_{B}=1,\ldots,M;\;
s_{1},s_{2}=\uparrow,\downarrow
\right\}
\label{superblockbasis}
\end{equation}

We write down the Hamiltonian
${\cal H}\tsub{S}\equiv{\cal H}_{A}+{\cal H}_{A\bullet}+
{\cal H}_{\bullet\bullet}+{\cal H}_{\bullet B}+{\cal H}_{B}$ for the 
superblock viz
\begin{eqnarray}
\bra{n'\tsub{S}}{\cal H}_{A}\ket{n\tsub{S}}
& = &
\delta_{s'_{1}s_{1}}\delta_{s'_{2}s_{2}}\delta_{n'_{B}n_{B}}
H(n'_{A},n_{A})
\label{superblockmatrixelements1}
\\
\bra{n'\tsub{S}}{\cal H}_{A\bullet}\ket{n\tsub{S}}
& = &
\delta_{s'_{2}s_{2}}\delta_{n'_{B}n_{B}}
\left[
S(n'_{A},n_{A})\bra{s'_{1}}S^{-}_{\bullet}\ket{s_{1}}+
\mbox{ h.c.}
\right]
\label{superblockmatrixelements2}
\\
& \vdots &
\nonumber
\end{eqnarray}
Low energy eigenstates of the 4-site Hamiltonian ${\cal H}\tsub{S}$ may 
be found at this point.

Next, we expand the system block by augmenting it with one of the extra 
sites. A basis for the augmented block $\tilde{A}\equiv A\bullet$ is
given by
\begin{equation}
\left\{
\ketd{n}
\equiv
\ket{n_{A}}\otimes\ket{s_{1}}
:n_{A}=1,\ldots,M;\;
s_{1}=\uparrow,\downarrow
\right\}
\label{augmentedblockbasis}
\end{equation}
and clearly has $\tilde{M}\equiv 2M$ states.

In order to again form a (larger, by two sites) superblock, we require 
the matrix elements of the augmented block Hamiltonian
${\cal H}_{\tilde{A}}\equiv{\cal H}_{A}+{\cal H}_{A\bullet}$ as well as 
the spin operator corresponding to the added (end) site viz
\begin{eqnarray}
\tilde{H}(n',n)
& \equiv &
\brad{n'}{\cal H}_{\tilde{A}}\ketd{n}
\\
& = &
\delta_{s'_{1}s_{1}}H(n'_{A},n_{A})+
\left[
S(n'_{A},n_{A})\bra{s'_{1}}S^{-}_{\bullet}\ket{s_{1}}+
\mbox{ h.c.}
\right]
\label{augmentedblockmatrixelements}
\\
\tilde{S}(n',n)
& \equiv &
\brad{n'}S^{+}_{\bullet}\ketd{n}
\\
& = &
\delta_{n'_{A}n_{A}}\bra{s'_{1}}S^{+}_{\bullet}\ket{s_{1}}
\end{eqnarray}

The superblock construction \Ref{superblockbasis}, 
\Ref{superblockmatrixelements1} and \Ref{superblockmatrixelements2} can 
now be extended indefinitely where $\tilde{A}\mapsto A$, 
$\tilde{H}\mapsto H$, $\tilde{S}\mapsto S$, 
$\tilde{M}\mapsto M$ and $\ketd{.}\mapsto\ket{.}$. However, even 
with the aid of sparse matrix diagonalisation methods, the size of the 
superblock basis, $O(M^{2})$ will soon grow prohibitively large as $M$ 
is doubled at each iteration.

Instead, we truncate the basis of the augmented block so that the number 
of states used in the subsequent iteration does not exceed a fixed 
threshold, $m$. The truncated basis is chosen so as to contain the 
{\em most important} states for forming low energy eigenstates of the 
${\cal H}\tsub{S}$. White's most important innovation was the use of a 
{\em density matrix} to determine these states.

For instance, if we are interested in the ground state 
$\ket{\psi\tsub{G}}$, then we form a reduced density matrix $\rho$ for 
the augmented block as follows
\begin{equation}
\brad{n'}\rho\ketd{n}
\equiv
\sum_{s_{2},n_{B}}
\langle n'\tsub{S}|\psi\tsub{G}\rangle
\langle\psi\tsub{G}|n\tsub{S}\rangle
\end{equation}
with
\[
\ket{n\tsub{S}}
\equiv
\ket{n_{A}}\otimes\ket{s_{1}}\otimes\ket{s_{2}}\otimes\ket{n_{B}}
\equiv
\ketd{n}\otimes\ket{s_{2}}\otimes\ket{n_{B}}
\]
\[
\ket{n'\tsub{S}}
\equiv
\ket{n'_{A}}\otimes\ket{s'_{1}}\otimes\ket{s_{2}}\otimes\ket{n_{B}}
\equiv
\ketd{n'}\otimes\ket{s_{2}}\otimes\ket{n_{B}}
\]
That is, the superblock ground state projection operator is traced over the 
environmental degrees of freedom $\bullet B$.

The density matrix eigenstates $\{\kett{n}:n=1,\ldots,\tilde{M}\}$ 
are found by standard dense matrix diagonalisation methods and the $m$ most  
important states (corresponding to the $m$ largest eigenvalues of 
$\rho$) are retained in forming the system block for the next iteration. 
Of course, the operators of interest must be represented in terms of 
these states viz
\begin{equation}
\tilde{H}(n',n)
\mapsto 
\sum_{n''n'''}
\langle\langle\langle
n'|n''
\rangle\rangle
\tilde{H}(n'',n''')
\langle\langle
n'''|n
\rangle\rangle\rangle
\mapsto
H(n',n)
\end{equation}
where $1\leq n,n'\leq\min\{\tilde{M},m\}\mapsto M$.

The density matrix eigenvalues sum to unity and the {\em truncation 
error}, defined as the sum of the density matrix eigenvalues corresponding 
to discarded 
eigenvectors gives a qualitative indication as to the accuracy of the 
calculation as well as providing a framework for extrapolation to the 
$m=\infty$ limit \cite{s=1_chain}. As a result of basis truncation, the 
cpu time is linear in the superblock size and so one can easily proceed 
to the thermodynamic limit for intrinsic properties such as the ground 
state energy density
(see \cite{thermodynamic_limit} for a formal discussion of the DMRG fixed 
point and the thermodynamic limit of the DMRG).

As mentioned, the accuracy of results obtained in this way was 
unprecedented \cite{s=1_chain}, the accuracy of the ground state energy 
density $e_{0}$ for the spin-1 chain being ultimately limited by the 
precision 
of machine arithmetic viz $e_{0} \cong -1.401484038971(4)$. Similarly 
exquisite accuracy persists when targeting low-lying excitation energies 
in various symmetry sectors 
\cite{s=1_chain} e.g.\ $E_{1}-E_{0}\cong 0.41050(2)$.

The method described above is the {\em infinite lattice algorithm}. In 
the {\em finite lattice algorithm}, the infinite lattice algorithm is 
initially applied so as to obtain a superblock of a desired size. System 
blocks are retained at all iterations and are used as environment blocks 
for the purpose of recursively deriving progressively superior system 
blocks within the superblock whose size remains fixed 
\cite{development}. Repeated {\em sweeps} across the lattice result in 
substantial improvements in accuracy, especially for the study of edge 
states and systems with impurities.

Applications to spin chains with high spin \cite{s=2}, \cite{s=32}, 
dimerisation and/or frustration \cite{bursill1}--\cite{pati}, together with 
extensions to coupled spin chains \cite{coupled_spin_chains}, models with 
itinerant fermions \cite{tJ_model}, \cite{hubbard_model}, Kondo systems
\cite{kondo_insulators}--\cite{kondo_necklace}, as well as coupled fermion 
chains \cite{fermion_ladders} have followed, even in cases of 
intermediate doping. Formulations for systems with single 
\cite{single_impurity} as well as randomly distributed 
\cite{random_impurities} impurities and disorder \cite{disorder} have also 
been forthcoming.

By retaining matrix elements of operators acting on sites throughout the 
blocks, it is possible to calculate static correlation functions. In 
\cite{s=1_chain}
and \cite{hallberg1} it is shown that highly accurate correlation 
functions can be calculated over many lattice spacings. Highly accurate 
studies of the structure factor and string order parameter (topological 
long range order) \cite{s=1_chain}, as well as edge states 
\cite{edge_states} in Haldane phase systems have also been performed. 
Dynamical correlation functions have also been calculated within the static 
\cite{static}, continued fraction \cite{hallberg2} and correction vector 
\cite{correction_vector} approaches. Finally, the DMRG has been 
formulated for the solution of models of spin chains {\em dynamically 
coupled} to (dispersionless) phonons \cite{phonons}.

\section{Spin chains with frustration}
\label{frustration}

A spin model on a bipartite lattice with a nearest neighbour, 
antiferromagnetic Heisenberg interaction can have its tendency towards 
commensurate order in the ground state {\em frustrated} in the presence 
of other interactions. In this section we consider two such models in 
one dimension. In each case, $J_{1}>0$ denotes the strength of the 
Heisenberg interaction and $J_{2}\geq 0$ the strength of the frustrating 
term.

The Majumda-Gosh (MG) model \cite{majumda_gosh} has the Heisenberg 
interaction frustrated by the presence of next-neighbour interactions viz
\begin{equation}
H = \sum_{i}
\left[
J_{1} S_{i}.S_{i+1} + J_{2}^{(MG)} S_{i}.S_{i+2}
\right]
\label{majumda_gosh}
\end{equation}
where $S_{i}$ denotes the spin $S=1/2$ operator for lattice site $i$. 
This model has some relevance to the interpretation of experiments on 
quasi one dimensional spin-1/2 chains \cite{emery}, \cite{zigzag}.
For  $S=1$, the order can also be disturbed in the Lai-Sutherland (LS)  
\cite{lai_sutherland} model through the presence of a biquadratic term
\begin{equation}
H = \sum_{i}
\left[
J_{1} S_{i}.S_{i+1} + J_{2}^{(LS)} (S_{i}.S_{i+1})^{2}
\right]
\label{lai_sutherland}
\end{equation}

The models \Ref{majumda_gosh} \cite{majumda_gosh}--\cite{zigzag}, 
\cite{nomura} and 
\Ref{lai_sutherland} \cite{lai_sutherland},
\cite{fath_solyom}--\cite{xiang_gehring} have been the subject of intense 
study in the past. Exact diagonalisation methods together with exact 
solutions and rigorous results at special points have been combined to give 
a good picture of their critical properties.

In the MG case, the pure Heisenberg point $(J_{2}=0)$ has been 
extensively studied \cite{heisenberg} and is a gapless spin liquid with 
quasi long range order (algebraically decaying correlations). This 
behaviour persists up to a critical point \cite{nomura} 
$J\tsub{2c}\cong 0.241167 J_{1}$ where a
Kosterlitz-Thouless (KT) transition occurs to a gapped, dimer phase with 
exponentially decaying correlations. A simple dimer wavefunction becomes 
exact at the MG point $(J_{2}=J_{1}/2)$ \cite{majumda_gosh}. Here it can 
be shown that the gap is non-zero \cite{affleck}. At 
$J_{2}=\infty$, the lattice separates into two uncoupled Heisenberg 
chains and is thus again critical.

For the LS model, the Heisenberg point is not critical, there being a 
non-zero gap and exponentially decaying correlations \cite{haldane}, 
\cite{s=1_chain}. At the Affleck point 
$(J_{2}=J_{1}/3)$ a simple valance bond solid (VBS) wavefunction 
provides the exact ground state \cite{affleck} and the model is 
rigorously known to be gapped \cite{affleck}. The model is 
solvable at the LS point $(J_{2}=J_{1})$ \cite{lai_sutherland}, having 
soft mode excitations at $k=\pm 2\pi/3$. The best numerical studies 
\cite{fath_solyom} indicate that this point is critical (i.e.\ 
$J\tsub{2c}=J_{1}$), separating a gapped VBS (Haldane) phase from a 
gapless phase with a quasi long ranged 3-fold ground state periodicity 
which persists up to the ferromagnetic boundary $(J_{2}=\infty)$, the 
transition again being of the KT type.

Until the advent of the DMRG, less was understood regarding the nature 
of incommensurate correlations in these models. In the classical limit 
$S=\infty$, these models both exhibit a crossover to incommensurate 
behaviour at some threshold $J\tsub{2cl}$. For the MG model we have 
$J\tsub{2cl}=J_{1}/4$ and $J\tsub{2cl}=J_{1}/2$ for the LS model. For 
$J_{2}\leq J\tsub{2cl}$ correlations are commensurate, spins being 
antialigned in a N\'{e}el ground state. For $J_{2}>J\tsub{2cl}$ the {\em 
classical pitch angle} $\theta\tsub{cl}$ between successive spins is 
given by
$\theta\tsub{cl}\equiv\cos^{-1}(-J\tsub{2cl}/J_{2})$. The DMRG has 
allowed a characterisation of the incommensurate behaviour through 
the accurate calculation of correlations over large distances within 
large 
chains.

In a series of studies of the structure factor
\begin{equation}
S(k)\equiv\frac{1}{N}\sum_{j,j'=1}^{N}C_{jj'}e^{i(j-j')k}
\end{equation}
where 
\begin{equation}
C_{jj'}\equiv\bra{\psi\tsub{G}}S_{j}^{z}S_{j'}^{z}\ket{\psi\tsub{G}}
\end{equation}
is the correlation function and $N$ the lattice size, a threshold for 
the onset of incommensurate correlations $J\tsub{2L}$ was defined as the 
value of $J_{2}$ at which the peak in the structure factor $k\tsub{max}$ 
begins to deviate from $\pi$ \cite{bursill1}--\cite{chitra}. Plots of 
$k\tsub{max}$ as well as $\theta\tsub{cl}$ for the two models are given in 
Fig.\ 1.

The DMRG permits accurate pinpointing of the threshold. For the MG model 
we have $J\tsub{2L}\approx 0.5206 J_{1}$ \cite{bursill2} and for the LS 
model $J\tsub{2L}\approx 0.4684 J_{1}$ \cite{bursill1}. Schollw\"{o}ck 
et al \cite{schollwock} studied $C(r)$, the correlation in real space. 
They observed the existence of a {\em disorder point} $J\tsub{2D}$ 
marking the onset of incommensurability in {\em real space}. They 
interpreted $J\tsub{2L}$ as being a {\em Lifshitz} point. In fact, it 
was shown for the LS model that $J\tsub{2D}=J_{1}/3$ i.e.\ the disorder 
point was precisely the Affleck point. Bursill et al \cite{bursill2} 
showed that for the MG model, $J\tsub{2D}=J_{1}/2$ i.e.\ the disorder 
point coincides with the MG point. $J\tsub{2D}$ lies in the gapped 
phase, near to where the gap in maximal \cite{schollwock}, 
\cite{zigzag}. Numerically it appears as though the correlation length 
$\xi$ takes on its minimum at $J_{2}=J\tsub{2D}$. Schollw\"{o}ck et al 
have pointed out that $J\tsub{2cl}$, $J\tsub{2L}$ and $J\tsub{2D}$ 
should be expected to merge in the classical limit $S\rightarrow\infty$.

\section{The DMRG and critical phenomena}
\label{critical}

A {\em critical point} in a quantum lattice model is a point in the 
parameter space at which the ground state energy is non-analytic. 
Generally, the energy gap $\Delta$ (between the ground and first excited 
states) vanishes as the critical point is approached and, 
correspondingly, the correlation length $\xi$ diverges.

As a result of the long range correlations in the vicinity of critical 
points, attempts to characterise them using exact diagonalisation 
methods may be hindered by limitations on the system sizes that can be 
reached. In principle, the DMRG can study very large systems with high 
accuracy and so it is natural to apply the DMRG to the study of critical 
points. In this section we review attempts to do so in magnetic models. 
We distinguish between methods which characterise critical points via 
direct calculations of $\Delta$ or $\xi$, and those which do so by 
generating renormalisation group transformations in the space of 
coupling constants.

\subsection*{Direct calculations of $\Delta$ or $\xi$}

The direct calculation of $\Delta$ or $\xi$ in the vicinity of a 
critical point within the DMRG is hindered by the fact that the 
convergence of calculated eigenvalues or correlation functions with $m$ 
is slowest when $\Delta$ is small or $\xi$ is large \cite{development}, 
\cite{bursill1}, \cite{schollwock}. Nevertheless, such calculations 
have been forthcoming.

Chitra et al \cite{chitra} have calculated $\Delta$ for the MG model 
\Ref{majumda_gosh} and $\xi$ \cite{bursill1}, \cite{schollwock}, and 
$\Delta$ \cite{schollwock} have been calculated for the LS model 
\Ref{lai_sutherland} but 
in all cases, though high accuracy is possible near the disorder point 
$J_{2}=J\tsub{2D}$, results become too inaccurate as the critical point 
$J\tsub{2c}$ is approached for any recovery of the finite size scaling 
results of \cite{nomura} and \cite{fath_solyom}.

For the MG model, Bursill et al \cite{bursill2} have also defined a {\em 
dimer order parameter}
\begin{equation}
D\equiv
\lim_{ N \rightarrow \infty }
\left| C_{ N/2 - 1 \; N/2 }-C_{ N/2 \; N/2 + 1 } \right|
\end{equation}
which characterizes the dimer phase $J\tsub{2c}<J_{2}<\infty$, vanishing 
as $J_{2}\rightarrow \left(J\tsub{2c}\right)^{+}$. A plot of $D$ versus 
$J_{2}/J_{1}$, calculated with $m=200$ is given in 
Fig.\ 2. Again, the critical point is not 
pinpointed with this approach.

These somewhat
disappointing results are not surprising given that the phase 
transitions are of the KT type, from a gapped phase to an extended 
gapless phase with an essentially singular divergence in $\xi$ which is 
extremely difficult to recover numerically without recourse to scaling 
theory.

For transitions such as the Ising transition where the order parameter 
vanishes only at the critical point and the system is gapped elsewhere, 
there is more hope for direct approaches. The exactly solvable $S=1/2$ 
transverse Ising (ITF) model
\begin{equation}
{\cal H} = -\gamma \sum_{i} S_{i}^{z} - \sum_{i} S_{i}^{x} S_{i+1}^{x}
\label{transverseisingmodel}
\end{equation}
has an Ising transition at $\gamma\tsub{c}=2$ which is accurately recovered 
by the direct approach \cite{bursill4}.

Kato and Tanaka \cite{kato_tanaka} have used the DMRG to pinpoint such a 
transition in the dimerised $S=1$ chain
\begin{equation}
{\cal H} = \sum_{i} ( 1 - (-1)^{i} \delta ) S_{i}.S_{i+1}
\end{equation}
at $\delta = 0.25 \pm 0.01$. Also, Pati et al \cite{pati} appear 
to have discovered such a point in the $S=1$ case of \Ref{majumda_gosh} 
at $ J_{2} = 0.730 \pm 0.05 $. Ladder models consisiting of two coupled 
Heisenberg chains have been studied \cite{coupled_spin_chains}. The phase 
transition whereby a gap opens on the introduction of interchain exchange, 
has been characterised.

Finally, by implementing new efficiency measures into the finite lattice 
algorithm, White and Affleck \cite{zigzag} have performed calculations on 
the MG model \Ref{majumda_gosh} for lattices with many thousands of sites 
using large values of $m$. These calulations afford an accurate 
study of $k\tsub{max}$, $\xi$ and $\Delta$ for relatively large values of 
$J_{2}$. With the help of field theory, this allows a characterisation of 
the critical point at $J_{2}=\infty$.

\subsection*{Coupling constant transformations from the DMRG}

As mentioned, critical points and exponents can be calculated by generating 
RG transformations in the space of coupling constants. This approach has 
been applied extensively using the RSRG to derive the transformations 
\cite{pfeuty}.

The idea is that a system of $N$ sites, with Hamiltonian $ {\cal H} $ 
paramterised by coupling constant(s) $\gamma$, is divided into $N/b$ blocks 
of $b$ sites. $N/b$-fold tensor products of the $m$ lowest energy 
eigenstates
\begin{equation}
B \equiv \{ \ket{\psi_{i}} : i=1,\ldots,m \}
\end{equation}
of the block Hamiltonians are used to form a truncated basis
$ B \otimes \ldots \otimes B $ (of dimension $ m^{N/b} $) for the system. 
$B$ is then identified as being a basis for a small repeat unit in a 
lattice model---a single site, or a {\em multisite}. For example, if $m=2$ 
then $B$ is identified as the basis for a site containing an $S=1/2$ 
operator, if $m=4$ then identification with a multisite consisting of two 
$S=1/2$ atoms is appropriate. $ {\cal H}' $, the projection of $ {\cal H} $ 
onto $ B \otimes \ldots \otimes B $ can then be identified with a 
spin Hamiltonian.

If $ {\cal H}' $ is of the same form as $ {\cal H} $, but with a different 
constant (vector) $\gamma'$ then
\begin{equation}
\gamma\mapsto\gamma' = T(b|\gamma)
\label{renormalisationgrouptransformation}
\end{equation}
constitutes a RG transformation from which critical points
\begin{equation}
\gamma\tsub{c} = T(b|\gamma\tsub{c})
\end{equation}
and critical exponents such as the thermal exponent
\begin{equation}
\nu = \frac{ \log b }{ \log | T'(b|\gamma\tsub{c}) | }
\end{equation}
can be extracted.

The RSRG results for the ITF model were quite acceptable \cite{sarker}. 
However, for the anisotropic $S=1/2$ Heisenberg ($XXZ$) model
\begin{equation}
{\cal H} = \sum_{i}
\left[
S_{i}^{z}S_{i+1}^{z}+
\frac{\gamma}{2}
\left( S_{i}^{+}S_{i+1}^{-} + \mbox{ h.c.} \right)
\right]
\label{anisotropicheisenbergmodel}
\end{equation}
which is essentially singular $ ( \nu = \infty ) $ at the Heisenberg 
point $ ( \gamma\tsub{c} = 1 ) $, the RSRG result was incorrect 
\cite{spronken} in that $ \nu \downarrow 2 $ as $ b \rightarrow \infty $.

It is natural to apply the DMRG philosophy to generate coupling constant 
transformations. That is, instead of using the $m$ lowest eigenstates of 
the block Hamiltonian to form $B$, the $m$ most probable states of a block 
immersed in an enviroment are used \cite{drzewinski}. This lead to 
dissapointing results for the ITF model, it being concluded that the DMRG 
offered no special advantage over the RSRG when applied to critical 
phenomena.

The studies in \cite{drzewinski} can not, however, be considered exhaustive 
as the block sizes used were quite small. In \cite{bursill3}, it was 
proposed that the robustness, portability and systematic improvability of 
the DMRG could be applied to the generation of RG transformations 
\Ref{renormalisationgrouptransformation} with large blocksizes $b$, using 
the fact that accurate representations of important states can be gleaned 
for 
large blocks.

This approach was applied to the $XXZ$ model. The result for $\nu$, given 
as a function of $b$ in Fig.\ 3, shows a marked 
improvement over the RSRG result, the essential singularity being recovered 
in the limit $ b \rightarrow \infty $ \cite{bursill3}. It would seem that 
more testing of the DMRG as a generator of RG transformations might be 
worthwhile.

Finally, it was noted in \cite{bursill3} that the values of the important 
eigenvalues in the various symmetry sectors of the density matrix provide a 
general and intuitive scheme for choosing a suitable value of $m$ and hence 
for identifying the renormalised Hamiltonian ${\cal H}'$. For instance, it 
was noted that the $ S=1 $ Heisenberg model mapped naturally onto a ladder 
model under this scheme, as has been demonstrated analytically and 
numerically \cite{white_mapping}.

\section{Recent extensions}
\label{current}

In this section we discuss some current trends in advancing the 
applicability of the DMRG. We focus on recent extensions of the DMRG to the 
calculation of thermodynamical properties of 2D classical systems and 
(hence) 1D quantum systems, as well as to the calculation of low energy 
properties of quantum models in 2D.

\subsection*{2D classical and 1D quantum systems at non-zero temperature} 

In a recent paper, Nishino exploited the similarities between the transfer 
matrix of a 2D classical system and the Hamiltonian of a 1D quantum system 
in order to formulate the DMRG for 2D classical systems \cite{nishino}. 
Results for the specific heat of the 2D Ising model are most promising, the 
exact result being recovered with good accuracy over the whole temperature 
range, even at criticality, using a reasonably modest basis set 
\cite{nishino}.

Potential applications of this approach include the determination of the 
critical properties of 2D classical models with high spin, anisotropy and 
frustration. Also, this extension paves the way for the study of quantum 
spin chains at non-zero temperature, as the Trotter-Suzuki decomposition 
allows such systems to be mapped onto classical 2D systems.

This programme has been carried out for the dimerised $XY$ chain
\begin{equation}
{\cal H}
=
-\sum_{i}
( S_{2i-1}^{x} S_{2i}^{x} + S_{2i-1}^{y} S_{2i}^{y} )
-\gamma\sum_{i}
( S_{2i}^{x} S_{2i+1}^{x} + S_{2i}^{y} S_{2i+1}^{y} )
\end{equation}
Although technical difficulties with the transfer matrix limit the basis 
sizes which can be used, very reasonable results, down to low temperatures 
$T$ have been obtained for the internal energy $u$, especially for 
$\gamma\neq 1$, when the model possesses a substantial gap \cite{bursill5}. 
A plot of $u$ versus $T$, for	$\gamma=2$ is given in Fig.\ 4.

	Finally, Nishino and Okunishi have derived two further 
reformulations of the DMRG---the product wavefunction renormalisation group 
(PWFRG) \cite{PWFRG} and the corner transfer matrix renormalisation group 
(CTMRG) \cite{CTMRG}. These formulations may offer dramatically improved 
efficiency over the DMRG and the means of calculating dynamical correlation 
functions in spin chains \cite{CTMRG}. The computational efficiency of the 
CTMRG has been exploited so as to obtain highly accurate results for the 2D 
Ising model at criticality \cite{CTMRG}.

\subsection*{Two dimensional quantum systems}

Soon after the inception of the DMRG, attempts were made at generalising 
the finite lattice algorithm to two dimensions
\cite{liang1}--\cite{liang2}. However, various conditions which permit the 
unprecedented accuracy of the DMRG in 1D---the fact that, for short ranged 
interactions there is no interaction between system and environment blocks, 
as well as the preservation of reflection symmetry, do not persist in 
higher dimensions. This is reflected in the relatively slow convergence of 
the DMRG when applied without modification to two dimensional systems 
\cite{liang1}.

In order to circumvent this problem, the DMRG has been reformulated for 
fermion systems in terms of momentum space basis states \cite{xiang1}. This 
advancement is made possible through an elegant solution to the technical 
problem of storing an updating the potential energy operator which is a sum 
of products of four creation and annihilation operators. The formulation 
offers the following distinct advantages:
\begin{enumerate}
\item
The total momentum can be used as a quantum number in diagonalising the 
superblock Hamiltonian.
\item
Rotational symmetry is not broken. The breaking of this symmetry is a major 
cause of the slow convergence of the real space formulation of the DMRG.
\item
A single program structure applies quite generally in all dimensions.
\end{enumerate}

Results for the 2D Hubbard model are extremely promising \cite{xiang1}. 
Agreement with exact calculations on $4 \times 4$ lattices is good as is 
agreement with projected quantum Monte Carlo (PQMC) and stochastic 
diagonalisation (SQ) data for $8 \times 8$ lattices. In fact, the DMRG, 
being a truncated basis expansion, generates a variational bound for the 
ground state energy. This bound is better than that obtained from SQ and is 
thus the best variational bound on the ground state energy to date (see 
table 1). Also, a numerical solution of the Hubbard model on a
$12 \times 12$ lattice has been afforded by this scheme \cite{xiang1}.

\begin{table}[htbp]
\centering

\begin{tabular}{|c|c|c|}
\hline
$N$ &   DMRG  & SQ$\tsup{a}$ \\
\hline
10  & -34.325 &   -34.31   \\
\hline
18  & -54.394 &   -54.37   \\
\hline
26  & -66.098 &   -66.05   \\
\hline
\end{tabular}
\\
{\footnotesize a) Ref.\ \protect\cite{SQ}.}

\caption{Comparison of DMRG and SQ results for the ground state energy of
the 2D Hubbard model on an $8 \times 8$ lattice, with $U=4t$ in the usual
notation, for various electron fillings $N$.}

\end{table}

Potential applications of this approach include heavily doped fermion 
systems, fermion systems with weak coupling and fermion systems with long 
range coulomb interactions.

Unfortunately, spin hamiltonians or models with dressed fermion operators 
such as the $t$-$J$ model cannot be conveniently expressed in terms of 
momentum space operators. However, there have been algorithmic advancements 
in the DMRG which have been exploited to permit the study of such models in 
2D \cite{CaV409}--\cite{xiang2}. White has performed successful 
calculations of the spin gap in a (frustrated) 2D Heisenberg model for the 
spin-gapped system CaV$_{4}$O$_{9}$ using a $24 \times 11$ lattice 
\cite{CaV409} and White and Scalapino have investigated the nature of one 
and two hole ground states in the $t$-$J$ model on $10 \times 7$ clusters
\cite{2DtJ}.

These calculations have been made possible by the use of highly efficient 
code which minimizes the number of hamiltonian matrix multiplications 
required in the sparse diagonalisation. Moreover, the problem of symmetry 
breaking which leads to the need for a large number of matrix 
multiplications has been recently addressed \cite{xiang2}.

In \cite{xiang2} an optimal scheme for {\em sweeping} the lattice has been 
developed. This has afforded the study of a frustrated Heisenberg model 
(i.e.\ with next-neighbour exchange) on a $20 \times 20$ lattice. Good 
convergence along with the accurate recovery of exact results on the
$4 \times 4$ lattice permits confident characterizations of the phases for 
various values of the coupling constants.

\subsubsection*{Acknowledgement}

R.\ J.\ B.\ gratefully acknowledges the support of SERC grant no.\ 
GR/J26748.

\section*{Figure cations}

\begin{enumerate}

\item
DMRG results for the structure factor peak position $k\tsub{max}$ and 
classical pitch angle $\theta\tsub{cl}$ (in units of $\pi$) as functions of 
the frustration $J_{1}/J_{2}$ for two frustrated spin systems---the MG and 
LS models.

\item
DMRG results for the dimer order parameter $D$ as a function of the 
frustration $J_{1}/J_{2}$ for the MG model. The known position of the
phase transition is indicated by the dashed line.

\item
DMRG result for the critical exponent $\nu$ as a function of blocksize $b$ 
for the anisotropic Heisenberg model. The dashed line is the result from 
the RSRG approach.

\item
DMRG result (dashed line) for the internal energy as a function of 
temperature for the dimerised $XY$ model with dimerisation ratio 
$\gamma=2$. The full line is the exact result.

\end{enumerate}


\begin{thebibliography}{99}

\bibitem{haldane}
F.\ D.\ M.\ Haldane, Phys.\ Rev.\ Lett.\ 50 (1983) 1153; Phys.\ Lett.\ 93A 
(1983) 464.

\bibitem{wilson}
K.\ Wilson, Rev.\ Mod.\ Phys.\ 47 (1975) 773.

\bibitem{classical}
Th.\ Niemeijer and J.\ M.\ J.\ van Leeuwen, Phys.\ Rev.\ Lett.\ 31 (1973) 
1411;
Th.\ Niemeijer and J.\ M.\ J.\ van Leeuwen, Physica 71 (1973) 17.

\bibitem{bray_chui}
J.\ W.\ Bray and S.\ T.\ Chui, Phys.\ Rev.\ B 19 (1979) 4876.

\bibitem{development}
S.\ R.\ White and R.\ M.\ Noack, Phys.\ Rev.\ Lett.\ 68 (1992) 3487;
S.\ R.\ White, Phys.\ Rev.\ Lett.\ 69 (1992) 2863,
Phys.\ Rev.\ B 48 (1993) 10 345.

\bibitem{s=1_chain}
S.\ R.\ White and D.\ A.\ Huse, Phys.\ Rev.\ B 48 (1993) 3844;
E.\ S.\ S\o rensen and I.\ Affleck, Phys.\ Rev.\ Lett.\ 71 (1993) 1633;
E.\ S.\ S\o rensen and I.\ Affleck, Phys.\ Rev.\ B, 49 (1994) 13235;
E.\ S.\ S\o rensen and I.\ Affleck, Phys.\ Rev.\ B, 49 (1994) 15771.

\bibitem{thermodynamic_limit}
S.\ \"{O}stlund and S.\ Rommer, Phys.\ Rev.\ Lett.\ 75 (1995) 3537.

\bibitem{s=2}
U.\ Schollw\"{o}ck and Th.\ Jolicoeur, Europhys.\ Lett.\ 30 (1995) 493;
Y.\ Nishiyama, K.\ Totsuka, N.\ Hatano and M.\ Suzuki, J.\ Phys.\ Soc.\ 
Jap.\ 64 (1995) 414.

\bibitem{s=32}
K.\ Hallberg, X.\ Q.\ G.\ Wang, P.\ Horsch and A.\ Moreo, {\em Critical 
behavior of the $S=3/2$ antiferromagnetic Heisenberg chain}, Preprint.

\bibitem{bursill1}
R.\ J.\ Bursill, T.\ Xiang and G.\ A.\ Gehring,  J.\ Phys.\ A 28 (1994) 
2109.

\bibitem{schollwock}
U.\ Schollw\"{o}ck, Th.\ Jolicoeur and T.\ Garel, Phys.\ Rev.\ B 53 
(1996) 3304.

\bibitem{bursill2}
R.\ J.\ Bursill, G.\ A.\ Gehring, D.\ J.\ J.\ Farnell, J.\ B.\ 
Parkinson, Tao Xiang and Chen Zeng, J.\ Phys.\ C 7 (1995) 8605.

\bibitem{chitra}
R.\ Chitra, S.\ Pati, H.\ R.\ Krishnamurthy, D.\ Sen and S.\ Ramasesha, 
Phys.\ Rev.\ B 52 (1995) 6581.

\bibitem{kato_tanaka}
Y.\ Kato and A.\ Tanaka, J.\ Phys.\ Soc.\ Jap.\ 63 (1994) 1277.

\bibitem{pati}
S.\ Pati, R.\ Chitra, D.\ Sen, H.\ R.\ Krishnamurthy and S.\ Ramasesha, 
Europhys.\ Lett.\ 33 (1996) 707.

\bibitem{coupled_spin_chains}
M.\ Azzouz, L.\ Chen and S.\ Moukouri, Phys.\ Rev.\ B 50 (1994) 6223;
S.\ R.\ White, R.\ M.\ Noack and D.\ J.\ Scalapino, 
Phys.\ Rev.\ Lett.\ 73 (1994) 886;
K.\ Hida, J.\ Phys.\ Soc.\ Jap.\ 64 (1995) 4896;
T.\ Narushima, T.\ Nakamura and S.\ Takada, J.\ Phys.\ Soc.\ Jap.\ 64 
(1995) 4322;
U.\ Schollw\"{o}ck and D.\ Ko, Phys.\ Rev.\ B 53 (1996) 240.

\bibitem{tJ_model}
L.\ Chen and S.\ Moukouri, Phys.\ Rev.\ B 53 (1996) 1866;
S.\ Moukouri, L.\ Chen and L.\ G.\ Caron, Phys.\ Rev.\ B 53 (1996) R488.

\bibitem{hubbard_model} 
S.\ J.\ Qin, S.\ D.\ Liang, Z.\ B.\ Su and L.\ Yu, Phys.\ Rev.\ B 52 (1995) 
R5475.

\bibitem{kondo_insulators}
C.\ C.\ Yu and S.\ R.\ White, Phys.\ Rev.\ Lett.\ 71 (1993) 3866;
C.\ C.\ Yu and S.\ R.\ White, Physica B 199 (1994) 454
M.\ Guerrero and C.\ C.\ Yu, Phys.\ Rev.\ B 51 (1995) 10301.

\bibitem{kondo_lattice}
S.\ Moukouri and L.\ G.\ Caron, Phys.\ Rev.\ B 52 (1995) 15723;
N.\ Shibata, T.\ Nishino, K.\ Veda and C.\ Ishii, Phys.\ Rev.\ B 53 
(1996) R8828.

\bibitem{anderson_lattice}
M.\ Guerrero and R.\ M.\ Noack, Phys.\ Rev.\ B 53 (1996) 3707.

\bibitem{kondo_necklace}
H.\ Otsuka and T.\ Nishino, Phys.\ Rev.\ B 52 (1995) 15066;
S.\ Moukouri, L.\ G.\ Caron, C.\ Bourbonnais and L.\ Hubert, Phys.\ 
Rev.\ B 51 (1995) 15920.

\bibitem{fermion_ladders}
R.\ M.\ Noack, S.\ R.\ White and D.\ J.\ Scalapino,
Phys.\ Rev.\ Lett.\ 73 (1994) 882;
S.\ R.\ White, R.\ M.\ Noack and D.\ J.\ Scalapino, J.\ Low Temp.\ 
Phys.\ 99 (1995) 593;
R.\ M.\ Noack, S.\ R.\ White and D.\ J.\ Scalapino, Europhys.\ Lett.\ 30 
(1995) 163.
C.\ A.\ Hayward, D.\ Poilblanc, R.\ M.\ Noack, D.\ J.\ Scalapino and W.\ 
Hanke, Phys.\ Rev.\ Lett.\ 75 (1995) 926.

\bibitem{single_impurity}
T.\ A.\ Costi, P.\ Schmitteckert, J.\ Kroha and P.\ W\"olfle, Phys.\ Rev.\ 
Lett.\ 73 (1994) 1275;
S.\ Eggert and I.\ Affleck, Phys.\ Rev.\ Lett.\ 75 (1995) 934;
E.\ S.\ S\o rensen and I. Affleck, Phys.\ Rev.\ B 51 (1995) 16115;
X.\ Q.\ Wang and S.\ Mallwitz, Phys.\ Rev.\ B 53 (1996) R492;
W.\ Wang, S.\ J.\ Qin, Z.\ Y.\ Lu, L. Yu and Z.\ B.\ Su, Phys.\ Rev.\ B 53 
(1996) 40; 
C.\ C.\ Yu and M.\ Guerrero, {\em An Anderson impurity in a 
semiconductor}, Preprint.

\bibitem{random_impurities}
K.\ Hida, J.\ Phys.\ Soc.\ Jap.\ 65 (1996) 895.

\bibitem{disorder}
P.\ Schmitteckert and U.\ Eckern,
Phys.\ Rev.\ B 53 (1996) 15397.

\bibitem{hallberg1} 
K.\ A.\ Hallberg, P.\ Horsch and G.\ Martinez, Phys.\ Rev.\ B 52 (1995) 
R719.

\bibitem{edge_states}
S.\ J.\ Qin, T.\ K.\ Ng and Z.\ B.\ Su, Phys.\ Rev.\ B 52 (1995) 12844.

\bibitem{static}
H.\ B.\ Pang, H.\ Akhlaghpour and M.\ Jarrell, Phys.\ Rev.\ B 53 (1996) 
5086.

\bibitem{hallberg2} 
K.\ A.\ Hallberg, Phys.\ Rev.\ B 52 (1995) R9827.

\bibitem{correction_vector}
S.\ K.\ Pati, S.\ Ramasesha, Z.\ Shuai and J.\ L.\ Bredas {\em Dynamical 
nonlinear optic coefficients from the symmetrized density matrix 
renormalization group method}, Preprint.

\bibitem{phonons}
L.\ G.\ Caron and S.\ Moukouri, Phys.\ Rev.\ Lett.\ 76 (1996) 4050.

\bibitem{majumda_gosh}
K.\ Majumda and D.\ K.\ Ghosh, 1969a, J.\ Math.\ Phys.\ 10 (1969) 1388;
{\em ibid} 1399.

\bibitem{emery}
G.\ Castilla, S.\ Chakravarty and V.\ J.\ Emery, Phys.\ Rev.\ Lett.\ 75 
(1995) 1823.

\bibitem{zigzag}
S.\ R.\ White and I.\ Affleck, {\em Dimerization and incommensurate spiral 
spin correlations in the zigzag spin chain: Analogies to the Kondo 
lattice}, Preprint.

\bibitem{lai_sutherland}
C.\ K.\ Lai, J.\ Math.\ Phys.\ 15 (1974) 1675;
B.\ Sutherland, Phys.\ Rev.\ B 12 (1975) 3795.

\bibitem{nomura}
K.\ Okamoto and K.\ Nomura, Phys.\ Lett.\ A 169 (1992) 433;
K.\ Nomura and K.\ Okamoto, J.\ Phys.\ Soc.\ Japan 62 (1993) 1123;
K.\ Nomura and K.\ Okamoto, J.\ Phys.\ A 27 (1994) 5773;
S.\ Eggert, {\em Numerical evidence for multiplicative logarithmic 
corrections from marginal operators}, Preprint.

\bibitem{fath_solyom}
G.\ Fath and J.\ Solyom, Phys.\ Rev.\ B 47 (1993) 872.

\bibitem{xiang_gehring}
T.\ Xiang and G.\ A.\ Gehring, J.\ Mag.\ and Mag.\ Mat.\ 104-107 (1992) 
861;
T.\ Xiang and G.\ A.\ Gehring, Phys.\ Rev.\ B 48 (1993) 303.

\bibitem{heisenberg}
E.\ Lieb and D.\ Mattis (editors), {\em The Many Body Problem---An 
Encyclopedia of Exactly Solvable Models in One Dimension} (1993).

\bibitem{affleck}
I.\ Affleck, T.\ Kenedy, E.\ H.\ Lieb and H.\ Tasaki, Phys.\ Rev.\ Lett.\ 
59 (1987) 799.

\bibitem{bursill4}
R.\ J.\ Bursill, F.\ Gode and G.\ A.\ Gehring, In preparation.


\bibitem{pfeuty}
P.\ Pfeuty, R.\ Jullien and K.\ A.\ Penson, in {\em Topics in Current 
Physics 30}, edited by T.\ W.\ Burkhardt and J.\ M.\ J.\ van Leeuwen 
(Springer-Verlag, Berlin, 1982).

\bibitem{sarker}
S.\ K.\ Sarker, Phys.\ Rev.\ B 30 (1984) 2752.

\bibitem{spronken}
G.\ Spronken, R.\ Jullien and M.\ Avignon, Phys.\ Rev.\ B 24 (1981) 
5356.

\bibitem{drzewinski}
A.\ Drzewi\'{n}ski and J.\ M.\ J.\ van Leeuwen, Phys.\ Rev.\ B 49 (1994) 
403;
A.\ Drzewi\'{n}ski and F.\ Daerden, J.\ Mag.\ and Mag.\ Mat.\ 140-4 
(1995) 
1623;
A.\ Drzewi\'{n}ski and R.\ Dekeyser, Phys.\ Rev.\ B 51 (1995) 15 218.

\bibitem{bursill3}
R.\ J.\ Bursill and F.\ Gode, J.\ Phys.\ C 7 (1995) 9765.

\bibitem{white_mapping}
S.\ R.\ White, Phys.\ Rev.\ B 53 (1996) 52.

\bibitem{nishino}
T.\ Nishino, J.\ Phys.\ Soc.\ Jap.\ 64 (1995) 3598;
T.\ Nishino, K.\ Okunishi and M.\ Kikuchi, Phys.\ Lett.\ A, 213 (1996) 
69.

\bibitem{bursill5}
R.\ J.\ Bursill, T.\ Xiang and G.\ A.\ Gehring,
J.\ Phys.\ C, 8 (1996) L1.

\bibitem{PWFRG}
T. Nishino and K.\ Okunishi, J.\ Phys.\ Soc.\ Jap.\ 64 (1995) 4084.

\bibitem{CTMRG}
T. Nishino and K.\ Okunishi, J.\ Phys.\ Soc.\ Jap.\ 65 (1996) 891.

\bibitem{liang1}
S.\ D.\ Liang and H.\ B.\ Pang, Phys.\ Rev.\ B 49 (1994) 9214.

\bibitem{liang2} 
S.\ Liang and H.\ Pang, Europhys.\ Lett.\ 32 (1995) 173.

\bibitem{xiang1}
T.\ Xiang, Phys.\ Rev.\ B 53 (1996) 10445.

\bibitem{SQ}
H.\ Deraedt and M.\ Frick, Phys.\ Rep.\ 231 (1993) 107.

\bibitem{CaV409}
S.\ R.\ White, {\em Spin Gaps in a Frustrated Heisenberg model for 
CaV$_4$O$_9$}, Preprint.

\bibitem{2DtJ}
S.\ R.\ White and D.\ J.\ Scalapino, {\em Hole and Pair Structures in 
the $t$-$J$ model}, Preprint.

\bibitem{xiang2}
T.\ Xiang, {\em Phase diagram of the frustrated Heisenberg model in
two dimensions}, In preparation.

\end{thebibliography}
\end{document}